\documentclass[twocolumn,aps,pra,floatfix,showpacs]{revtex4}

\usepackage{graphicx,color}
\usepackage{dcolumn}
\usepackage{bm}
\usepackage{mathbbol}
\usepackage{amsmath,amssymb}  
\usepackage[latin9]{inputenc} 

\makeatletter

\newtheorem{prop}{Proposition}
\newtheorem{defin}{Definition}
\newtheorem{thm}{Theorem}
\newtheorem{cor}{Corollary}
\newtheorem{lemma}{Lemma}

\newcommand{\proof}{\noindent {\bf Proof. }}

\newcommand{\ket}[1]{|#1\rangle}
\newcommand{\bra}[1]{\langle #1|}

\newcommand{\Hi}{\mathcal{H}}

\newcommand{\Ei}{\mathcal{E}}

\newcommand{\supp}{\textrm{supp}}

\newcommand{\beq}{\begin{equation}}
\newcommand{\eeq}{\end{equation}}
\newcommand{\bea}{\begin{eqnarray}}
\newcommand{\eea}{\end{eqnarray}}
\newcommand{\beqan}{\begin{eqnarray*}}
\newcommand{\eeqan}{\end{eqnarray*}}

\newcommand{\R}{\mathbb{R}}
\newcommand{\cvd}{\hfill $\Box$ \vskip 2ex}

\newcommand{\tr}{\textrm{trace}}

\begin{document}

\title{Quantum Information Encoding, Protection, and Correction 
\\ from Trace-Norm Isometries}

\author{Francesco Ticozzi}
\email{ticozzi@dei.unipd.it}
\affiliation{Dipartimento di Ingegneria dell'Informazione,
Universit\`a di Padova, via Gradenigo 6/B, 35131 Padova, Italy}
\author{Lorenza Viola}
\email{lorenza.viola@dartmouth.edu}
\affiliation{\mbox{Department of Physics and Astronomy,
Dartmouth College, 6127 Wilder Laboratory, Hanover, NH 03755, USA}}

\date{\today}

\begin{abstract}
We introduce the notion of trace-norm isometric encoding and explore
its implications for passive and active methods to protect quantum
information against errors.  Beside providing an operational
foundations to the ``subsystems principle'' [E. Knill, Phys. Rev. A
{\bf 74}, 042301 (2006)] for faithfully realizing quantum information
in physical systems, our approach allows additional explicit
connections between noiseless, protectable, and correctable quantum
codes to be identified.  Robustness properties of isometric encodings
against imperfect initialization and/or deviations from the intended
error models are also analyzed.
\end{abstract}


\maketitle

\section{Introduction}

The idea that states of ideal quantum systems, carrying abstractly
defined quantum information, must be suitably mapped -- {\em encoded}
-- into states of a physical system, in such a way that information
can be best protected against the unavoidable effect of errors,
underpins the possibility to practically exploit the added power of
quantum information in real-world devices.  According to the so-called
{\em subsystems principle}
\cite{viola-qubit,viola-generalnoise,verification,knill-protected},
logically mapping quantum information into a subsystem of a Hilbert
space provides the most general approach to quantum encoding, and
well-identifiable subsystems must exist at each point in time in order
for the desired information to be faithfully represented throughout a
computational process.  Subsystem-encodings play a central role in the
theory of quantum fault tolerance, allowing, in particular, for a
unified understanding of quantum error control to be gained in terms
of passive protection based on decoherence-free subspaces
\cite{zanardi-DFS,lidar-DFS} and noiseless subsystems
\cite{viola-generalnoise}, as well as active stabilization based on
either ``initialization-protectable'' or ``error-correcting
subsystems'' \cite{knill-protected}.  Conceptually, the subsystems
principle provides the foundation for operator quantum error
correction (OQEC) \cite{kribs-QEC,kribs-OEC,Brun-EAQEC}, which is the
most general error-control framework presently known for noise
described by a completely positive trace-preserving (CPTP) map.

For physically realized information, recent work by Blume-Kohout and
coworkers \cite{viola-IPS,IPS-Long} has shown that preservation of the
{\em mutual distinguishability} between states under a given error
process is key to a general {\em operational} characterization of the
information-preserving structures (IPS) that the process can support,
whether in passive or active form.  Mathematically, the starting point
is to realize that preservation of information in a set of possible
states (a {\em code}) under the action of a map ${\mathcal E}$ is
equivalent to requiring that ${\mathcal E}$ acts on the code as a
distance-preserving map, where the appropriate distance measure is
induced by the {\em trace norm}.  Remarkably, since the latter yields
both lower and upper bound to the fidelity between quantum states
\cite{fuchs-distance}, the trace norm provides the appropriate metric
of performance for quantifying distance between open-system evolutions
\cite{lidar-distance,CDCG}.

In the light of the special significance that both subsystems and
trace-norm isometries have in the broad QEC context, a natural
question arises: Can these notions be related at a fundamental level?
Equivalently, {\em what role do trace-norm isometries play in
representing quantum information?}  Exploring the implications of a
description that directly exploits trace-norm isometries is the main
motivation of this work.  We show that by insisting in the requirement
that quantum information encodings be $1$-isometries, a number of {\em
a priori} unrelated results are consistently recovered, and additional
new insight is gained.  In particular, our first result (Sec. II) is
the possibility to {\em derive} a manifestation of the subsystems
principle, thereby firmly grounding it on operational requirements.
In Sec. III, an explicit form of the most general codes that can
faithfully encode quantum information and of the class of
transformations that preserve and recover these codes is obtained.  In
the process, we elucidate connections between the dual notions of {\em
correctability} and {\em protectability} that were not captured by the
previous IPS analysis \cite{viola-IPS}, and further characterize QEC
scenarios whereby the required active intervention may be achieved
through purely unitary means
\cite{spekkens-unitary,choi-multiplicative}.  In Sec. IV, we argue
that the trace-norm isometric approach developed for describing
perfect quantum information encoding and recovery may serve as a
useful starting point for investigating `perturbations' around exact
notions, thus complementing ongoing investigations of {\em
approximate} QEC \cite{Tyson,beny-approximate,beny-alg,Ng2009}.  We
conclude in Sec. V with some open questions.


\section{Quantum Information Encodings}

Consider an ideal quantum system ${\cal Q},$ defined on a Hilbert
space $\Hi_{\cal Q},$ with states belonging to the trace-one,
positive, convex subset ${\cal D}(\Hi_{\cal Q})$ of the Hermitian
operators ${\cal O}(\Hi_{\cal Q})$, representing physical
observables. ${\cal Q}$ represents the {\em logical} level, the
abstract quantum information to be encoded.  Our task is to represent
quantum information carried by ${\cal Q}$ in a given quantum {\em
physical} system ${\cal P}$, defined on a Hilbert space $\Hi_{\cal
P},$ with corresponding states ${\cal D}(\Hi_{\cal P})$ and
observables ${\cal O}(\Hi_{\cal P})$.

An {\em encoding} of ${\cal Q}$ in ${\cal P}$ is specified once two
maps, a {\em state encoding} $\Phi$ and an {\em observable encoding}
$\Psi$, are given:
\begin{eqnarray}
\label{statesubset}\Phi:{\cal D}(\Hi_{\cal Q})\rightarrow \Sigma ,
\\
\label{observablesubset}\Psi:{\cal O}(\Hi_{\cal Q})\rightarrow
\Omega,
\end{eqnarray}
where the elements of $\Sigma$ and $\Omega$ are nonempty sets of
${\cal D}(\Hi_{\cal P})$ and ${\cal O}(\Hi_{\cal P}),$
respectively. The use of subsets instead of single operators allows
for the possibility that the encoding is insensitive (robust) against
the choice of a `co-subsystem' state, as it will be clear shortly (see
discussion after Theorem 1).

\vspace*{1mm}

In Ref.  \onlinecite{knill-protected}, Knill has formalized the
meaning of a {\em faithful} encoding, by requiring that the encoding
maps satisfy three physically-motivated conditions as follows:

\vspace*{1mm}

(i) Statics: For all $\sigma\in\Phi(\rho)$, $X\in\Psi(A),$ expectation
values coincide on faithfully encoded states: $ \tr(\sigma X)=\tr(\rho
A);$

(ii) Unitary dynamics: For all $\sigma\in\Phi(\rho)$,
$X\in\Psi(A),$ $e^{-iX}\sigma e^{iX}=e^{-iA}\rho e^{iA};$

(iii) Measurement dynamics: For all $\sigma\in\Phi(\rho),$ $
X\in\Psi(A),$ with $X=\sum_\lambda \lambda\Pi_X^\lambda$,
$A=\sum_\alpha \alpha\Pi_A^\alpha$ denoting the corresponding spectral
representations \cite{Remark1}, projective measures are faithfully
implemented in the sense that $ \Pi_X^\lambda\sigma \Pi_X^\lambda\in
\Phi(\Pi_A^\lambda\rho \Pi_A^\lambda).$

\vspace*{1mm}

It is then proved (Thm. 1 in \cite{knill-protected}) that every {\em
faithful encoding of quantum information is a subsystem encoding},
that is, there exists a decomposition 
\beq
\Hi_{\cal P}=\Hi_S\otimes\Hi_F\oplus\Hi_R,
\label{subdec}
\eeq
\noindent 
such that for all $\rho$, $\supp(\Phi(\rho))\subset
\Hi_S\otimes\Hi_F,$ and for all $X\in \Psi(A)$, $\Psi(A)$ is of the
form $X_S\otimes I_F\oplus X_R.$
Motivated by the operational requirement of distinguishability
preservation between sets of states \cite{viola-IPS,IPS-Long}, we now
introduce {\em 1-isometric encodings}, and compare them with faithful
encodings. In what follows, we shall primarily focus on encoding of
states, which is the key for quantum information protection and
correction in the Schroedinger's picture.

It is well known \cite{holevo,petz-qstatistics, nielsen-chuang} that
the probability of correctly distinguishing a pair of quantum states
is related to the distance induced by the trace-norm,
$$\|A\|_1\equiv \tr(|A|) = \sum_i s_i(A),$$ where $|A|\equiv
\sqrt{A^\dagger A}$ and $s_i(A)$ are the singular values.
Specifically, let two states $\rho,\,\tau,$ be prepared with prior
probability $p,1-p,$ respectively.  Then they can be discriminated by
means of measurements with at most probability
$\frac{1}{2}(1+\|p\rho-(1-p)\tau\|_1)$.  This naturally prompts
investigating the structure of encodings that preserve
distinguishability:

\begin{defin} 
A linear map on Hermitian operators, $\Phi:{\cal
O(H_Q)}\rightarrow{\cal O(H_P)}$, defines a 1-isometric encoding if
for all $\rho_1,\rho_2\in{\cal D(H_Q)}$ and $p\in[0,1]$,
\begin{equation} 
\label{isom} 
\|p\Phi(\rho_1)-(1-p)\Phi(\rho_2)\|_1=\|p\rho_1-(1-p)\rho_2\|_1.  \eeq
\end{defin}
Notice that linearity is assumed here from the beginning, reflecting
the fact that, in practice, any physical procedure to be employed as
an information ``encoder'' can be described as a linear state
transformation.  While distinguishability preservation might at first
seem a weak requirement in comparison with (i)-(iii) for faithful
encodings, we shall next show that it is indeed enough to enforce a
subsystem structure. We begin with a preliminary Lemma:

\begin{lemma} 
If Eq. \eqref{isom} holds for any $\rho_1,\rho_2\in{\cal D(H_Q)}$, the
map $\Phi$ is a linear isometry on the whole ${\cal O(H_Q)}.$
\end{lemma}
\proof Any $Z\in{\cal O (H_Q)}$ may be expressed in the form 
$$Z= Z^+ - Z^-= \tr(Z^+)\rho_+ - \tr(Z^-)\rho_- ,$$ where $Z^{+,-}$
are the positive and negative parts of $Z,$ respectively, and
$\rho_{+,-}=Z^{+,-}/\tr(Z^{+,-})\in {\cal D(H_Q)}.$ By letting
$p={\tr(Z^+)}/{\tr(Z^+ + Z^-)}$, we get \beqan\|\Phi(Z)\|_1 &=&
\|\tr(Z^+)\Phi(\rho_+) - \tr(Z^-)\Phi(\rho_-)\|_1\\ &=&(\tr(Z^+
+Z^-))\|p\Phi(\rho_+) - (1-p)\Phi(\rho_-)\|_1\\ &=&(\tr(Z^+ +
Z^-))\|p\rho_+ - (1-p)\rho_-\|_1\\ &=&\|\tr(Z^+)\rho_+ -
\tr(Z^-)\rho_-\|_1\\ &=&\|Z\|_1.  \eeqan
\noindent
Thus, $\Phi$ is an isometry on $\cal O(H_Q).$ 
\cvd

Since $\Phi:{\cal O}(\Hi_{\cal Q})\rightarrow {\cal O}(\Hi_{\cal P})$
is a linear $1$-isometry that sends states to states, it defines a
{\em stochastic isometry} in the terminology of Busch
\cite{busch-isometries}.  By invoking Thm. 1 of
Ref. \onlinecite{busch-isometries}, in particular, it follows that for
every $1$-isometric encoding as defined above, there exists a
decomposition of the form
\begin{equation}
\label{Hdec}\Hi_{\cal P}=\Big( \bigoplus_j\Hi_{S,j}\Big)\oplus
\Hi_R,\eeq 
\noindent 
with each $\Hi_{S,j}$ isomorphic to $\Hi_Q,$ such that
\begin{equation}
\label{1isom}
\Phi(\rho)=\Big(\bigoplus_j\omega_jU_j\rho U_j^\dag\Big)\oplus
\hat0_R,\eeq 
\noindent 
where $U_j:\Hi_{\cal Q}\rightarrow\Hi_{S,j}$ is either unitary or
anti-unitary, $\omega_j\in[0,1],\,\sum_j\omega_j=1$, and $\hat0_R$
denotes the zero operator on $\Hi_R$ (on the topic of isometric
mappings between quantum states, see also
\cite{molnar-isometries,Remark2}).  
%
%
Thus, up to a unitary (or anti-unitary) transformation $U_{\cal
P}=\big(\bigoplus_j U^\dag_j\big)\oplus \hat I_R$ on $\Hi_{\cal P}$,
and a possible reordering of the basis, it follows that {\em for any
1-isometric state encoding there exists a subsystem decomposition} of
$\Hi_{\cal P}$ of the form given in Eq. (\ref{subdec}),
such that
\begin{equation}
\label{states} 
\Phi(\rho)=\rho\otimes \tau\oplus \hat 0_R,\eeq 
\noindent
with a density operator $\tau$ on the co-subsystem factor $\Hi_F$ with
spectrum $\{\omega_j\}.$

We remark that the subsystem decomposition of the Hilbert space
$\Hi_{\cal P}$ associated to a given encoding $\Phi$ is in general {\em
not} unique. In particular, there exists a {\em minimal} decomposition
for which the state $\tau$ in \eqref{states} is {\em full-rank} in
$\Hi_F$.
The other subsystem decompositions of $\Hi_{\cal P}$ may be obtained
from the minimal one by augmenting the dimension of $\Hi_F$ (thus
reducing the one of the summand $\Hi_R$) upon identifying more
isomorphic copies of $\Hi_{S,j}\sim\Hi_{\cal Q}$ in \eqref{Hdec},
associated to weights $\omega_j=0$ in \eqref{1isom}. The latter
subspaces do not actually encode any information, since the state has
trivial support there.

Once a subsystem decomposition of $\Hi_{\cal P}$ is chosen, a natural
observable encoding $\Psi$ is given by
\begin{equation}
\label{observables}
\Psi(A)=A\otimes I_F\oplus X_R,\eeq 
\noindent 
for some $X_R\in{\cal O}(\Hi_R)$ \cite{Remark3}. Given the structure
of the encoded states in Eq.  \eqref{states}, the specific choice of
non-minimal subsystem decomposition and of $X_R$ is irrelevant for
expectations, dynamics, and measurement on the encoded states. In
fact, one may directly verify that any pair $(\Phi,\Psi)$ of the form
given in Eqs. \eqref{states}-\eqref{observables} defines a faithful
encoding, and that the requirements (i)-(iii) do not depend on the
co-factor state, $\tau$.  Conversely, consider a faithful encoding
$(\bar\Phi,\bar\Psi).$ Then the associated subsystem structure
provides us with a class of 1-isometric state encodings $\Phi_\tau$ as
in Eq. \eqref{states}, parametrized by the state of the co-factor
$\tau\in{\cal D}(\Hi_F).$ Each pair $(\Phi_\tau,\bar\Psi)$ is a
faithful encoding.  We can summarize these properties in the
following:

\begin{thm} 
To every 1-isometric encoding $\Phi$ is associated a (minimal)
faithful subsystem encoding of the form given in
Eqs. \eqref{states}--\eqref{observables}, and to every faithful
encoding $(\bar\Phi,\bar\Psi)$ is associated a class of 1-isometric
encodings $\Phi_\tau$ parametrized by the co-factor state
$\tau\in{\cal D}(\Hi_F).$
\end{thm}

This result provides an explicit connection between 1-isometric and
faithful encodings. In fact, Thm. 1 may be regarded as establishing a
{\em subsystems principle} building on the operational notion of
distinguishability.

By requiring the state encoding $\Phi$ to be a {\em linear and
isometric} function, we lose in principle some of the structure
associated with the general encoding maps of \cite{knill-protected}
into {\em subsets},
Eqs. \eqref{statesubset}--\eqref{observablesubset}.  Nonetheless, it
is important to appreciate that to each subsystem decomposition is
associated a {\em class} of isometric encodings, parametrized by the
cofactor state $\tau$.  The latter are operationally indistinguishable
with respect to the faithfulness requirements (i)-(iii), as long as
they they share the same observable encoding of the form
\eqref{observables}. One can then think to describe such a class in a
compact form as a state encoding into subsets:
$$\hat\Phi:\rho\in{\cal D(H_Q)}\mapsto
\Sigma_{\rho}=\{\rho\otimes\tau\oplus\hat 0_R| \tau\in {\cal
D(H_F)}\}.$$
\noindent 
When the observable encoding is defined as in \eqref{observables},
$\hat\Phi$ can be interpreted as a {\em robust} encoding with respect
to $\tau$, in the sense that the desired information is correctly
represented in ${\mathcal H}_S$ irrespective of which element of
$\Sigma_\rho$ has been used.  Such a robustness property plays a
crucial role for characterizing the potential of error correction and
protection of $1$-isometric quantum codes, to which we turn next.


\section{Noiseless, Protectable, and Correctable Codes}

Consider a $1$-isometric encoding $(\Phi,\Psi)$ of ${\cal Q}$ in
${\cal P}$: We shall henceforth denote $\Phi({\cal D(H_Q)})$ by ${\cal
C}_{\cal Q}$ and call it a {\em code}.  Assume that as a result of
some noise process, the physical system ${\cal P}$ undergoes CPTP
dynamics described by a quantum operation ${\cal E}$ on ${\cal
D}(\Hi_{\cal P})$.  We start by recalling three desirable properties
that ${\cal C}_{\cal Q}$ may exhibit with respect to ${\cal E}$,
following the definitions given in Blume-Kohout {\em et al.}
\cite{viola-IPS}:

\begin{defin} 
A code $\cal C_{\cal Q}$ is (i) {\em fixed} by ${\cal E}$ if ${\cal
E}(\rho)=\rho$ for every $\rho\in \cal C_{\cal Q}$; (ii) {\em
preserved} by ${\cal E}$ if ${\cal E}$ acts as a 1-isometry on $\cal
C_{\cal Q}$; (iii) {\em noiseless} for ${\cal E}$ if it is preserved
by any convex mixture $\sum_k p_k{\cal E}^k$, with $p_k \geq0$ and
$\sum_k p_k =1$.
\label{preservednotions}
\end{defin}

Our focus in this Section is to study in detail how, within the
present $1$-isometric framework, the above properties relate to both
passive and active methods for stabilizing information encoded in
$\cal C_{\cal Q}$ against ${\cal E}$.

\subsection{Noiseless isometric codes and subsystems}
\label{subsystems}

The noiselessness property as stated in Definition
\ref{preservednotions} appears at first quite different from the
original concept underlying decoherence-free subspaces (DFSs)
\cite{zanardi-DFS,lidar-DFS} and, more generally, noiseless subsystems
(NSs)
\cite{viola-generalnoise,knill-protected,verification,kribs-oqec}.
While a number of equivalent characterizations exist, the following
may be taken as the standard defining property of a NS: Given a fixed
decomposition of the Hilbert space, $\Hi_{\cal
P}=\Hi_S\otimes\Hi_F\oplus\Hi_R$, and a TPCP $\cal E$, $\Hi_S$
supports a NS for ${\cal E}$ if for every $\rho_S \in {\cal
D}(\Hi_S)$, $\tau_F \in {\cal D}(\Hi_F)$, \beq\label{NS} {\cal
E}(\rho_S \otimes \tau_F) = \rho_S \otimes \sigma_F, \eeq
\noindent 
for some state $\sigma_F \in {\cal D}(\Hi_F)$.  That is, the
restriction of ${\cal E}$ to ${\Hi_S\otimes\Hi_F}$ obeys \beq
\label{NS1} 
{\cal E}|_{\Hi_S\otimes\Hi_F}=I_S \otimes {\cal F},\eeq
for some TPCP ${\cal F}$ on $\Hi_F.$ 

From Eq. \eqref{NS}, it is easy to show that a noiseless $1$-isometric
code exists with support on the same factor $\Hi_S$. In fact, it
suffices to consider a state $\tau \in{\cal D(H_F)}$ which is a fixed
point for ${\cal F},$ and observe that the code $\rho\otimes\tau
\oplus \hat0_R$ is fixed for ${\cal E},$ and hence it is trivially
noiseless. The converse is not equally straightforward: {\em Given
that a TPCP map admits a noiseless 1-isometric code, is there a
noiseless subsystem that shares the same (or a compatible) subsystem
structure?}

The rest of this section is devoted to prove that this is indeed the
case. We begin with the following Lemma:

\begin{lemma} 
Let ${\cal E}:{\cal D}(\Hi_{\cal P})\rightarrow {\cal D}(\Hi_{\cal
P})$ be a TPCP map and $\bar\rho\in{\cal D}(\Hi_{\cal P})$ such that
${\cal E}(\bar\rho)=\bar\sigma,$ with $\supp
(\bar\sigma)\subseteq\supp (\bar\rho).$ Let ${\cal D}(\bar \Hi)$ be
the set of density operators with support only on {\em
$\bar\Hi=\supp(\bar\rho).$} Then ${\cal E}({\cal D}(\bar\Hi))\subset
{\cal D}(\bar\Hi)$.
\label{invariance}
\end{lemma}

\proof Let us choose an operator-sum representation ${\cal
E}(\cdot)=\sum_k M_k \cdot M_k^\dag$.  Consider the orthogonal
decomposition $\Hi_{\cal P}=\bar\Hi\oplus\bar\Hi^\perp$: In a
block-matrix representation consistent with such a decomposition, we
may write
$$\bar\rho=
\left(
\begin{array}{c|c}
 \rho_S & 0  \\ \hline
 0 &    0 
\end{array}
\right),
\quad M_k=
\left(
\begin{array}{c|c}
 M_{k,S} & M_{k,P}  \\ \hline
 M_{k,Q} &    M_{k,R} 
\end{array}
\right),
$$
$${\cal E}(\bar\rho)=
\left(
\begin{array}{c|c}
 * & *  \\ \hline
 * &    \sum_k M_{Q,k} \rho_S M_{Q,k}^\dag
\end{array}
\right)=\bar\sigma,\quad
\bar\sigma=
\left(
\begin{array}{c|c}
 \sigma_S & 0  \\ \hline
 0 &  0 
\end{array}
\right). $$ 
\noindent 
Then it must be $\sum_k M_{Q,k} \rho_S M_{Q,k}^\dag=0,$ and since
$\rho_S$ is full-rank on $\bar\Hi,$ this implies $M_{Q,k}=0$ for all
$k.$ It is then easy to verify that this ensures ${\cal E}({\cal
D}(\bar\Hi))\subseteq {\cal D}(\bar\Hi).$ \cvd

By using the previous Lemma, we first show that if the action of the
map respects a fixed Hilbert space decomposition, the link between
noiseless isometric codes and NSs can be established directly:

\begin{thm}
\label{equalsupport} 
Consider a fixed Hilbert space decomposition $\Hi_{\cal
P}=\Hi_S\otimes\Hi_F\oplus\Hi_R$ and a TPCP map ${\cal E}:{\cal
D}(\Hi_{\cal P})\rightarrow {\cal D}(\Hi_{\cal P}).$ Assume that there
exists a full-rank state $\tau\in{\cal D}(\Hi_F)$, such that for every
$\rho\in{\cal D}(\Hi_S)$,
\begin{equation}
{\cal
E}(\rho\otimes\tau\oplus\hat0_R)=\rho\otimes\sigma\oplus\hat{0}_{R},\eeq
for some $\sigma\in{\cal D}(\Hi_{F}).$ Then
\begin{equation} 
{\cal E}|_{\Hi_S\otimes\Hi_F}={I}_S\otimes {\cal F}, \eeq
\noindent 
for some TPCP ${\cal F}:{\cal D}(\Hi_F)\rightarrow {\cal D}(\Hi_F).$
\end{thm}

\proof First consider a full-rank state $\rho\in{\cal D}(\Hi_S).$ Then
$$\supp(\rho\otimes\sigma\oplus\hat{0}_{R'})\subseteq \Hi_S
\otimes\Hi_F=\supp(\rho\otimes\tau\oplus\hat{0}_R),$$ and Lemma
\ref{invariance} implies that
$${\cal E}|_{\Hi_S\otimes\Hi_F}({\cal D}(\Hi_S\otimes\Hi_F))
\subseteq{\cal D}(\Hi_S\otimes\Hi_F).$$  
\noindent 
We can then restrict our attention to $\Hi_S\otimes\Hi_F.$ Consider an
operator-sum representation for ${\cal
E}|_{\Hi_S\otimes\Hi_F}(\cdot)=\sum_k M_k \cdot M_k^\dag.$ Let
$\{C_i\otimes D_j\}$ be an orthonormal operator basis for the
corresponding operator space ${\cal B}(\Hi_S\otimes\Hi_F),$ with
$C_0=I_S$.  We may thus rewrite
$$M_k=\sum_{i,j}m_{kij}C_i\otimes D_j= \sum_iC_i\otimes F_{ik},$$
 with $F_{ki}=\sum_j m_{kij}D_j.$  It then follows that
$${\cal E}(\rho\otimes\tau\oplus\hat0_R)= \sum_{klm}C_l\rho
C_m^\dag\otimes F_{lk}\tau F_{mk}^\dag=\rho\otimes \sigma.$$ 
\noindent 
Since the $\{C_i\}$ are orthonormal, it must be
$$\sum_{k}C_l\rho C_l^\dag\otimes F_{lk}\tau F_{lk}^\dag=0,$$ 
\noindent 
for $l>0$ and all $\rho\in{\cal D}(\Hi_S).$ This implies
$\sum_{k}F_{lk}\tau F_{lk}^\dag=0$ and, since $\tau$ is full rank by
hypothesis, $F_{lk}=0$ for $l\neq0.$ We thus get the conclusion with
${\cal F}(\cdot)=\sum_k F_{0k} \cdot F_{0k}^\dag.$ \cvd

In the general case where the Hilbert space decomposition is allowed
to change, the argument is less direct.  Building on the above
results, we have the following:

\begin{thm} 
Let ${\cal C}\sim\rho\,\otimes\sigma\oplus\hat 0_R$ be a $1$-isometric
noiseless for ${\cal E}$.  Then there exists a fixed code ${\cal
C}'\sim\rho\otimes\tau\oplus\hat 0_R$ that admits a subsystem
decomposition in common with ${\cal C}$ and for which, denoting by
$\Hi_{{\cal C}'}$ the support of ${\cal C'},$ ${\cal E}|_{\Hi_{{\cal
C}'}}={I}_S\otimes {\cal F}.$
\label{thm:NS}
\end{thm}

\proof Consider an infinite, convergent series
$$\bar\Ei=\sum_{i=0}^\infty p_i \Ei^i,\quad \sum_{i=0}^\infty
p_i=1,\,p_i > 0.$$ 
\noindent 
Then, any convex combination ${\cal G}=\sum_i q_i \Ei^i,$ with $\sum_i
q_i =1,$ $q_i\geq 0,$ satisfies $\supp({\cal G}({\cal C}))\subseteq
\supp(\bar{\Ei}({\cal C})).$ Define $\Hi_{SF}\equiv
\supp(\bar{\Ei}({\cal C}))$ and let $\rho\otimes\bar\tau\oplus\hat0_R$
be a minimal subsystem representation of $\bar\Ei({\cal C}).$ Consider
an orthonormal basis $\{\ket{\varphi_j}\}$ of $\Hi_{\cal Q}$, and the
associated orthogonal projections
$\Pi_{j}=\ket{\varphi_j}\bra{\varphi_j}.$ Define the isomorphic
subspaces $\Hi_{F,j}\equiv \supp(\Pi_j\otimes\bar\tau).$ By
hypothesis, the combination of the encoding and the evolution
$\bar\Ei\circ\Phi$ must act like a trace-norm isometry, and trace-norm
isometries preserve orthogonality \cite{busch-isometries}. Thus, by
definition of $\bar\Ei$ and orthogonality preservation, it follows
that one can define a family of orthogonal subspaces
\begin{eqnarray*}
\Hi_{F,j}&=&\supp(\bar\Ei(\Pi_j\otimes\sigma)), \;\;\forall j,\\
\Hi_{F,j}&\perp&\Hi_{F,k},\;j\neq k.
\end{eqnarray*}
\noindent 
Pick now an orthonormal basis in each $\Hi_{F,j},$ such that, for
instance, $\bar\Ei(\Pi_j\otimes\sigma)$ is diagonal in $\Hi_{F,j}$. We
can then construct a decomposition
\beq
\Hi_{SF}=\bigoplus_j \Hi_{F,j}=\Hi_S\otimes\Hi_F 
\label{SubS}
\end{equation}
\noindent 
Since, by definition of $\bar \Ei$, it must also be
$\supp(\Ei^i(\Pi_j\otimes\sigma))\subseteq \Hi_{F,j}$ for any $i,$ the
subsystem structure in (\ref{SubS}) supports not only $\bar\Ei({\cal
C}),$ but each $\Ei^i({\cal C})$ in subsystem form up to a unitary
change of basis in each of the $\Hi_{F,j}$'s. That is,
$$\Ei^i({\cal C})\sim U^{(i)}\Big(\rho\otimes\tau(i)\oplus\hat0_R
\Big)U^{(i)\dag}$$
\noindent 
on $\Hi_S\otimes\Hi_F\oplus\Hi_R,$ with $\Hi_R=\Hi\ominus \Hi_{SF}$
and a block-unitary $U^{(i)}=\bigoplus_j U_j^{(i)}\oplus I_R$ which in
general depends on the exponent $i$ of ${\cal E}^i$. The same
holds for any convex combination of powers of
${\cal E}$: In particular, consider the convex
combination
$$\Ei_N=\sum^N_{i=0} \frac{1}{N+1}\Ei^i.$$ Then the limit
$\Ei_\infty=\lim_{N\rightarrow\infty}\Ei_N$ is well defined for
finite-dimensional Hilbert spaces. In addition,
$\Ei\circ\Ei_{\infty}=\Ei_\infty,$ hence $\Ei_{\infty}$ projects onto
the fixed points of $\Ei$ \cite{viola-IPS,lindblad-nocloning}.
Furthermore, ${\cal C}_\infty \equiv \Ei_{\infty}({\cal C})$ must be
of the form $\rho\otimes\tau_\infty\oplus\hat0_R$ with respect to the
decomposition $\Hi_{\cal P}=\Hi_S\otimes\Hi_F\oplus\Hi_R,$ after some
change of basis $U_\infty=\Big(\bigoplus_j U_{\infty,j}\Big) \oplus
I_R,$ constructed as described above. It therefore follows that
$$\Ei(\rho\otimes\tau_\infty\oplus\hat0_R)=
\rho\otimes\tau_\infty\oplus\hat0_R,$$
\noindent 
for all $\rho\in{\cal D}(\Hi_{\cal P}).$ If we restrict to the support
of the code, we can apply Theorem \ref{equalsupport}, hence it follows
that $\Ei|_{\supp({\cal C}_\infty)}= I_S\otimes {\cal F}.$ \cvd

\subsection{Correctable vs. protectable codes} 

When no DFS or NS can be found under the error dynamics induced by
$\cal E$, active intervention via a recovery quantum operation
${\mathcal R}$ is required in order to protect quantum information
encoded in ${\mathcal P}$. For a given ${\mathcal R}$, the subsystems
principle implies that stored information must exist irrespective of
whether active intervention is effected {\em before} or {\em after}
error events take place, in a suitable sense
\cite{knill-protected,viola-generalnoise}. This is formalized by the
following:

\begin{defin}
A code $\cal C_{\cal Q}$ is (i) {\em correctable} for $\cal E$ if
there exists a CPTP map $\cal R$ on ${\cal D}(\Hi_{\cal P})$ such that
$\cal C_{\cal Q}$ is noiseless for ${\cal R}\circ{\cal E}$; (ii) {\em
protectable} for $\cal E$ if $\cal C_{\cal Q}$ is noiseless for ${\cal
E}\circ{\cal R}.$
\end{defin}

Two specializations of the above definitions will also be relevant: we
shall call {\em unitarily correctable (protectable)} a code for which
$\cal R$ can be chosen to be a unitary transformation on
$\Hi_{\mathcal P}$ (see also \cite{spekkens-unitary,IPS-Long}), and
{\em completely correctable (protectable)} a code which is {\em fixed}
for $\cal R \circ E$ (or, respectively, $\cal E \circ R$).

Note that what we refer to as completely correctable codes directly
correspond to the original notion of a (finite-distance) QEC code, in
which case the state of the syndrome subsystem (in the language of
\cite{knill-QEC}, Thm. III.5) has to be appropriately re-initialized
to its reference state at each iteration, and the code subspace
effectively forms a DFS under $\cal R \circ \cal E$ \cite{knill-QEC}.
NSs and OQEC, on the other hand, require from the outset that the
state of the syndrome subsystem be irrelevant as long as information
is properly encoded in the logical factor
\cite{viola-generalnoise,viola-IPS,kribs-oqec}. While OQEC does not
lead to fundamentally different quantum codes (in the sense that for
each OQEC code, an associated subspace QEC code may be found),
simplified recovery procedures may result from taking explicit
advantage of the subsystem structure \cite{Bacon}. In the framework of
$1$-isometric codes, our first result is to show how the
correspondence between NSs and noiseless isometric codes highlighted
in the previous section is complemented by the following
correctability property:

\begin{thm}
A $1$-isometric code ${\cal C}_{\cal Q}$ is preserved iff it is
correctable, and it is correctable iff it is completely correctable.
\label{correctability1}
\end{thm}

\proof Assume that ${\cal C}_{\cal Q,}$ of the form given in
Eq. \eqref{states}, is preserved by $\cal E.$ Hence ${\cal
E}\circ\Phi$ is a 1-isometry from ${\cal D}(\Hi_{\cal Q})$ on ${\cal
D}(\Hi_{\cal P})$ and, by Theorem 1, ${\cal E}({\cal C}_{\cal Q})$ must
correspond to another subsystem state-encoding of ${\cal Q}$ in ${\cal
P}$.  Assume the two subsystem decompositions of $\Hi_{\cal P},$
corresponding to ${\cal C}_{\cal Q}$ and ${\cal E}({\cal C}_{\cal
Q}),$ respectively, to be $\Hi_S\otimes\Hi_F\oplus\Hi_R$ and
$\Hi_{S'}\otimes\Hi_G\oplus\Hi_T,$ with
$\dim(\Hi_{S})=\dim(\Hi_{S'})=\dim(\Hi_{\cal Q}).$ The action of $\cal
E$ restricted to ${\cal C}_{\cal Q}$ can be represented in the form
$${\cal E}|_{\cal C_Q}={\cal U}_{S-S'}\otimes {\cal E}_{F-G},$$ 
\noindent 
with ${\cal U}_{S-S'}$ a unitary map from ${\cal D}(\Hi_S)$ to ${\cal
D}(\Hi_{S'}),$ and ${\cal E}_{F-G}$ a CPTP map from ${\cal D}(\Hi_F)$
to ${\cal D}(\Hi_G)$, respectively.  Let ${\cal
E}_{F-G}(\tau)=\sigma$.  Thus, given the definition of a noiseless
code given in Definition 2, ${\cal C_Q}$ is corrected by any $\cal R$
which obeys
$${\cal R}|_{\cal {\cal E}(C_Q)}={\cal U}^\dag_{S-S'}\otimes {\cal
R}_{G-F},$$ 
\noindent 
with ${\cal R}_{G-F}(\sigma)=\tau.$ One may for instance employ the
time-reversal of ${\cal E}_{F-G}$
\cite{barnum-reversing,ticozzi-htheorem}: That is, let $\{E_k\}$ be
the Kraus operators associated to ${\cal E}_{F-G},$ then its
time-reversal with respect to $\sigma$ is ${\cal R}_{{\cal
E}_{F-G},\sigma}$ with Kraus operators $\{\sigma^\frac{1}{2} E_k^\dag
\tau^{-\frac{1}{2}}\}.$ Notice that not only does this correction
operation make ${\cal C_Q}$ a noiseless code, but actually a code of
fixed points. 

Conversely, preservation is necessary for correctability since every
CPTP map acts like a trace-norm contraction on states (see {\em e.g.}
\cite{nielsen-chuang}):
\begin{equation}
\label{contraction} \|{\cal E}(\rho_1)-
{\cal E}(\rho_2)\|_1\leq \|\rho_1-\rho_2\|_1, \; \forall
\rho_1,\rho_2\in{\cal D}(\Hi_{\cal P}).\eeq 
\noindent 
If ${\cal C_Q}$ is not preserved by $\cal E,$ there must exist
$\rho_1,\rho_2\in{\cal C_Q}$ such that $ \|{\cal E}(\rho_1)- {\cal
E}(\rho_2)\|_1< \|\rho_1-\rho_2\|_1$.  To correct those states, any
correction $\cal R$ should violate \eqref{contraction}, and hence it
would not be a physically realizable TPCP map.  \cvd

Remarkably, a similar reasoning allows us to extend the analysis to
protectable codes:

\begin{cor}\label{protectable}
A TPCP map ${\cal E}$ admits a $1$-isometric protectable code ${\cal
C}_{\cal Q}$ iff it admits a $1$-isometric completely protectable code
${\cal C}'_{\cal Q},$ and it admits a $1$-isometric completely
protectable code ${\cal C}'_{\cal Q}$ iff it admits a $1$-isometric
completely correctable code ${\cal C}''_{\cal Q}$.
\end{cor}

\proof Clearly, if ${\cal C_Q}''$ is completely protectable it is also
protectable. It then suffices to prove that the existence of a
protectable code implies the existence of a completely correctable
one, and this in turn implies the existence of a completely
protectable one.  Assume that ${\cal C}_{\cal Q},$ of the form given
in Eq. \eqref{states}, is protectable for $\cal E,$ with a
``protecting'' quantum operation ${\cal R}$.  Then, by the above
argument, ${\cal C}_{\cal Q},$ ${\cal R}({\cal C}_{\cal Q})$, and
${\cal E}\circ{\cal R}({\cal C}_{\cal Q})$ must be $1$-isometric state
encodings of ${\cal Q}$ in ${\cal P}$ of the subsystem form.  Let
${\cal C}'_{\cal Q}={\cal R}({\cal C}_{\cal Q}).$ Then ${\cal C}'$ is
preserved by ${\cal E}$, hence completely correctable by the previous
Theorem.

Conversely, assume then that ${\cal C}'_{\cal Q}$ is completely
correctable for $\cal E,$ with correction operation ${\cal R'}$. Let
${\cal C}''_{\cal Q}={\cal E}({\cal C}'_{\cal Q}).$ Then we have
$$({\cal E}\circ{\cal R}') ({\cal C}''_{\cal Q})={\cal E}({\cal
C}'_{\cal Q})={\cal C}''_{\cal Q},$$
\noindent 
which is completely protectable.  \cvd 

The role of isometries for $1$-isometric subsystem encodings is
illustrated in Figure \ref{fig1}.  Note that while Theorem
\ref{correctability1} above may be regarded as a direct counterpart of
Theorem 1 in Ref. \onlinecite{viola-IPS}, the isometric approach has
the advantage of additionally providing the explicit structure of the
encoding, along with the explicit noise action and the required
correction map.  Furthermore, Corollary \ref{protectable} formally
establishes how protectable $1$-isometric encodings are in fact
equivalent to correctable ones.  Intuitively, at least as far as
perfect information recovery is concerned, we cannot hope to find a
protectable code if no correctable codes of the same dimension are
available.

\begin{figure}[t]
\begin{center}
\includegraphics[width=7.5cm]{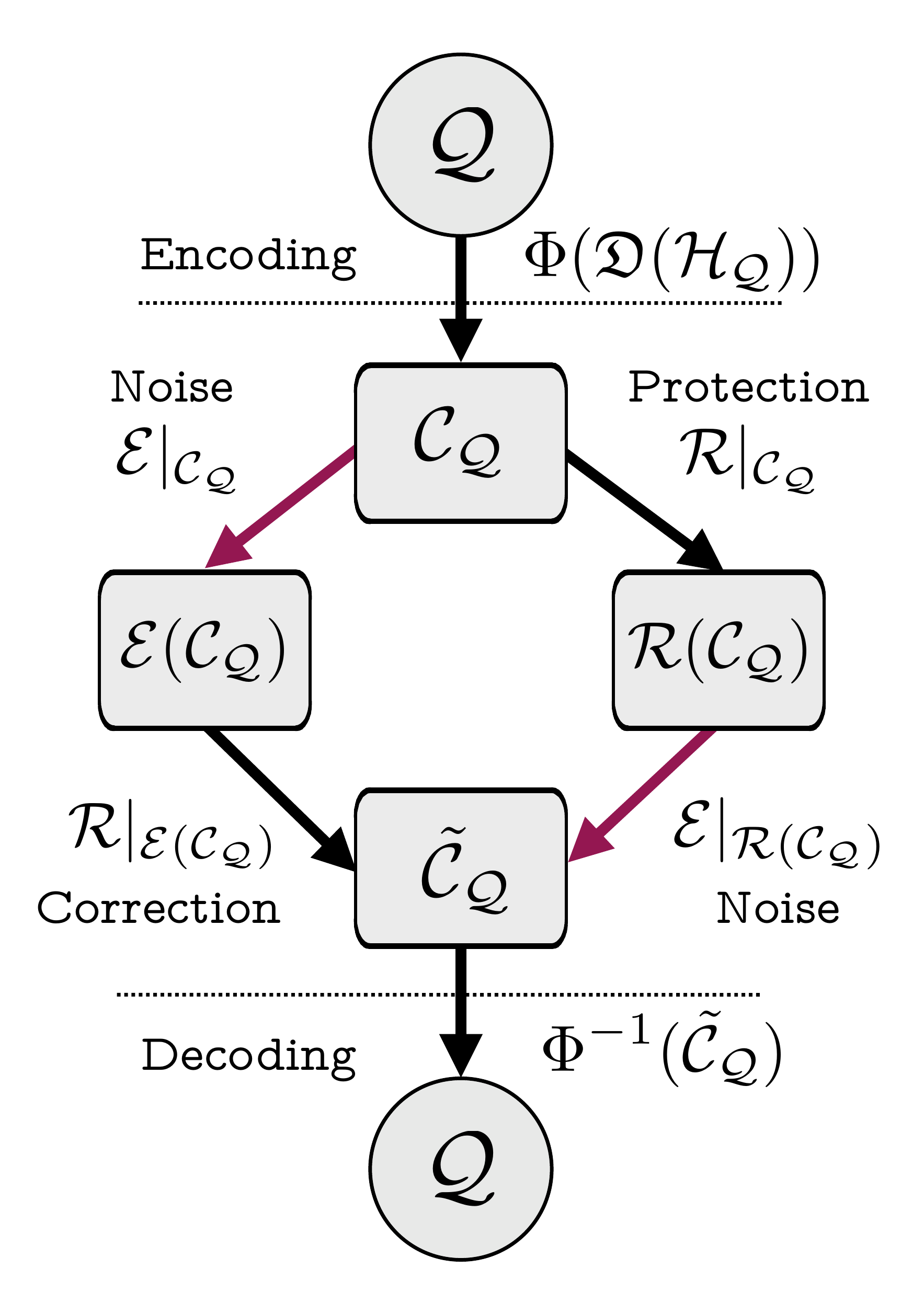}
\caption{(Color online) Pictorial representation of trace-norm
isometries in quantum information encoding, correction, and
protection. Here, $\tilde{{\cal C}}_Q$ denotes a code isomorphic to
${{\cal C}}_Q$, whereas the (state-)decoding map returns information
from the physical back to the abstract level.}
\label{fig1}
\end{center}
\end{figure}

Next, we consider the potential of unitary correction superoperators
for $1$-isometric codes. Our main result shows that unitary recovery
may suffice, provided that the code is not increasingly ``smeared''
under the noise action, in the following sense:

\begin{prop} 
Let $\cal C_Q$ be a preserved $1$-isometric code for $\cal E$, with
{\em$\textrm{dim}\left(\supp({\cal E
(C_Q)})\right)\leq\textrm{dim}({\cal\supp( C_Q}))$}. Then $\cal C_Q$
is unitarily correctable.
\end{prop}

\proof Consider a $1$-isometric code $\cal C_Q$ associated to
$\Hi_{S}\otimes\Hi_F\oplus\Hi_R$ and preserved by ${\cal E}.$ Thus,
there exists a minimal Hilbert space decomposition
$\Hi_{S'}\otimes\Hi_G\oplus\Hi_T$ associated to ${\cal E(C_Q)},$ with
$\dim(\Hi_{S})=\dim(\Hi_{S'}).$ If $\dim(\Hi_{G})<\dim(\Hi_{F}),$ we
may extend the decomposition to an non-minimal one
$\Hi_{S'}\otimes\Hi_{G'}\oplus\Hi_{T'},$ in such a way that equality
holds.  Having ensured that $\dim(\Hi_{G'})=\dim(\Hi_{F}),$ the two
subsystem representations are isomorphic, and there exists a unitary
correction superoperator ${\cal U}(\cdot)=U(\cdot)U^\dag$ that
restores the initial state up to the cofactor, which has rank at most
equal to the initial one.  Thus, Theorem \ref{equalsupport} applies to
$\cal U \circ E,$ and the code is supported on a NS for $\cal U \circ
E$. Therefore, it is unitarily correctable.  \cvd
  
When the main assumption of the above Theorem is violated, that is,
$\textrm{dim}(\supp({\cal E (C_Q)}))\leq\textrm{dim}({\cal\supp(
C_Q})),$ we can still gain some insight on what can be achieved by
restricting to unitary corrections: If, in the previous proof,
$\dim(\Hi_{G})>\dim(\Hi_{F}),$ we can define a non-minimal subsystem
encoding for ${\cal C_Q}$ such that $\dim(\Hi_{G})=\dim(\Hi_{F})$.
Then there exists a unitary change of basis $V$ such that
$V(\rho\otimes{\cal E_{F'-G'}}(\tau'))V^\dag$ is a 1-isometric
encoding for ${\cal Q}$ in $\Hi_{S}\otimes\Hi_{F'}\oplus\Hi_{R'},$
equivalent to the initial subsystem decomposition. Since the support
of ${\cal C_Q}$ is strictly contained in ${\cal E(C_Q)},$ Theorem
\ref{equalsupport} does not apply.  This means that every 1-isometric
subsystem encoding which is correctable is in general only {\em
unitarily recoverable}, that is, there exists a unitary operation that
restores the code to one which is supported by a non-minimal extension
of the initial subsystem decomposition. This is weaker than the code
being unitarily correctable, since there is {\em no} guarantee that
further iterations of noise and correction will still preserve the
code. We have thus recovered Theorem 1 in Ref.
\onlinecite{spekkens-unitary} (stating the equivalence between
correctable and unitarily recoverable subsystems) as a corollary of
our isometry-based analysis.  An interesting open question for further
investigation is to what extent direct connections might exist between
with the characterization of unitarily correctable codes from the
multiplicative domain of CPTP maps as recently pursued in
\cite{choi-multiplicative}.  Within the current analysis, we conclude
our discussion on perfect isometric encodings by considering a simple
illustrative example.

\vspace*{1mm}

\noindent
{\bf Example 1: The 3-bit quantum repetition code revisited}. Consider
a system of three qubits, described in $\Hi_{\cal
P}\approx(\mathbb{C}^2)^{\otimes 3}$ with the standard computational
basis
$\{\ket{abc}=\ket{a}\otimes\ket{b}\otimes\ket{c}\,|\,_{a,b,c\in\{0,1\}}\}$.
Assume that the dominant noise on the system stems from independent
bit-flip errors with probability $p< 1/2$, that is, ${\cal E}
\sim\{\sqrt{1-p}\, \sigma_0^{(i)},
\sqrt{\frac{p}{3}}\,\sigma_x^{(i)}\}_{i=1,2,3},$ where
$\sigma_x^{(i)}$ are Pauli operators on the $i$-th qubit, and
$\sigma_0^{(i)}=I$ corresponds to no-error.  Consider the subspace
$\Hi_{\cal C}=\textrm{span}\{\ket{000},\ket{111}\},$ which is used to
encode the states of a logical qubit ${\cal Q}$, that is, in our
notation, ${\cal C}={\cal D}(\Hi_C).$ Let us consider the subsystem decomposition $\Hi_{\cal P} \sim \Hi_{\cal Q} \otimes \Hi_F,$ induced by the unitary
change of basis $U$ defined by
\begin{equation}
U\ket{abc}=\ket{x}\otimes \ket{yz},
\label{rep}
\eeq 
\noindent 
where $x$ is the majority count of the binary string $abc,$ and $yz$
indicates in which location $abc$ differs from $xxx$, with $00$
indicating no differences.  Then, in the subsystem representation
defined by Eq. (\ref{rep}), $\Hi_{\cal C} \sim \Hi_{\cal Q} \otimes
|00\rangle$, ${\cal C}= \{ \rho\otimes \ket{00}\bra{00}\},$ with
$\rho\in{\cal D(H_Q)}$, see also \cite{viola-qubit}.  It is easy to see
that the action of $\cal E$ {\em restricted to $\cal C$} in this
representation is ${\cal E}|_{\cal C}=I\otimes {\cal F}$, where
explicitly
\begin{eqnarray*}
{\cal F}(\ket{00}\bra{00})\equiv \sigma
&\hspace*{-1mm}=\hspace*{-1mm}& (1-p) \ket{00}\bra{00} \\
&\hspace*{-1mm}+\hspace*{-1mm}& \frac{p}{3}(\ket{01}\bra{01}+
\ket{10}\bra{10}+\ket{11}\bra{11}).
\end{eqnarray*}

Thus, ${\cal E(C)}=\{\rho\otimes\sigma\}$ is $1$-isometric to ${\cal
C}$.  By using a correction superoperator ${\cal R}=I\otimes {\cal
A},$ where for instance ${\cal A}$ is a two-qubit amplitude damping
channel on $\Hi_F$ (such that for any $\sigma,$ ${\cal
A}(\sigma)=\ket{00}\bra{00}),$ it follows that ${\cal C}$ is fixed for
${\cal R\circ E}$.  Equivalently, $\Hi_{\cal C}$ is a DFS under ${\cal
R\circ E}$.
On the other hand, one may directly verify that the code ${\cal
C}'\equiv {\cal E(C)}$ is completely protectable for $\cal E$, with
protection operation $\cal R$ as above, as it follows from the proof
of Corollary \ref{protectable}.  Equivalently, $\Hi_{\cal Q}$ supports
a NS of ${\cal E\circ R}$.  

While the fact that {\em every} (finite-distance) subspace QEC code is
associated to a NS of ${\cal E\circ R}$ is implied by Theorem 6 in
\cite{viola-generalnoise}, OQEC subsystem codes take explicit
advantage of the fact that a NS may also be identified for ${\cal
R\circ E}$, for the same ${\cal R}$, see \cite{Bacon} for the simplest
representative of an OQEC code which protects one logical qubit into
nine qubits subject to arbitrary independent single-qubit errors.

\vspace*{1mm}


\section{Robustness of Isometric Encodings}

The $1$-isometric setting we analyzed thus far naturally lends itself
to investigate robustness properties of subsystem encodings.  In turn,
this allows to place our approach in the context of approximate QEC,
which has been extensively investigated for subspace codes from
different perspectives (see {\em e.g.}  \cite{barnum-reversing,
schumacher-approximate, klesse-approximate, buscemi-approximate,
leung-approximate}), and is now receiving renewed attention in the
light of extensions to the more general case of OQEC and subsystem
codes \cite{beny-approximate,beny-alg,Ng2009}.  While a comprehensive
study is beyond our current scope, we focus here on two representative
kinds of uncertainty sources:

\begin{itemize}
\item[$\bullet$] {\em Encoding uncertainty:} The procedure that
implements the intended encoding map may be affected by errors.

\item[$\bullet$] {\em Model uncertainty:} The noise model for which
the code and correction operator are intended may be only
approximately known.
\end{itemize}

For $1$-isometric encodings as defined in Sec. I, we show here that
error correction is robust with respect to bounded encoding errors.
Furthermore, an upper bound may be given on how information is
degraded in the presence of modeling errors. The key idea is to invoke
perturbations of $1$-isometries:

\begin{defin} 
A map $\tilde\Phi:{\cal O(H_A)}\rightarrow {\cal O(H_B)}$ is a
perturbation of a $1$-isometry if there exist a $1$-isometry
$\Phi:\cal O(H_A)\rightarrow {\cal O(H_B)}$ and $\varepsilon\in\R^+$
such that 
$$  \| \tilde\Phi (\rho) - \Phi (\rho) \|_1\leq\varepsilon,\;\;\;
\forall \rho\in{\cal D}(\Hi_{\cal Q}). $$
\end{defin}

Approximate isometries on Banach spaces have been introduced by Hyers
and Ulam in the 40's \cite{hyers-ulam} in their most general form
\cite{Remark4}, and since then their properties have been studied
intensively, most notably approximations with linear or affine
isometries as well as various extension problems, see {\em e.g.}
\cite{bhatia-semrl,dilworth,rassias,liu-approximate} and references
therein.  Note that in our context we consider only approximate
isometries that are {\em $\varepsilon$-perturbations of linear
1-isometries} and, for the sake of brevity, we still refer to them
simply as $\varepsilon$-isometries.  The definitions of approximate
$1$-isometric encodings and of approximately preserved codes are then
the natural extension of the exact ones to this setting.  In
particular:

\begin{defin}
A state encoding $\tilde\Phi$ of ${\cal Q}$ in ${\cal P}$ is an
$\varepsilon$-isometric encoding if $\tilde\Phi$ is a
$\varepsilon$-isometry.
\end{defin}

It follows from our previous analysis that $\varepsilon$-isometric
state encodings have the following structure:
\begin{equation}
\label{approxstates}
\tilde\Phi(\rho)=\rho\otimes \tau\oplus \hat 0_R +\Delta(\rho),\eeq
\noindent 
with $\|\Delta(\rho)\|_1\leq \varepsilon$.  In this way, given the
properties of the trace-norm, and the relation of the latter to
distinguishability, the perturbation parameter $\varepsilon$ inherits
the role of an {\em upper bound} to the error probability in
information recovery via measurements. The other relevant definitions
are also extended in a similar fashion:

\begin{defin}
A code {${\cal C_Q}$ is (i) {\em $\varepsilon$-preserved} for $\cal E$
if ${\cal E|_{C_Q}}$ is a $\varepsilon$-isometry; (ii) {\em
$\varepsilon$-noiseless} for $\cal E$ it is $\varepsilon$-preserved
for any convex combination of the form $\sum_k p_k{\cal E}^k$; (iii)
{\em $\varepsilon$-correctable} if it is $\varepsilon$-noiseless for
${\cal R}\circ{\cal E}$.}
\end{defin}

\subsection{Encoding uncertainty} \label{encunc}

The correction operations for $1$-isometric codes we discussed in
Sec. III.B exhibit a desirable property: Bounded errors in the
encoding map do not increase. This is formalized by the following:

\begin{prop}
Let $\Phi$ be a $1$-isometric encoding such that its range ${\cal
C_Q}$ is preserved by ${\cal E}$.  Then any $\varepsilon$-approximate
version $\tilde\Phi$ generates $\varepsilon$-approximate codes
$\tilde{\cal C}_{\cal Q}$ which are $\varepsilon$-correctable for
${\cal E}.$
\end{prop}

\proof Under the assumptions, any $\tilde{\cal C}_{\cal Q}$ is of the
form
$$\tilde\Phi(\rho)=\rho\otimes \tau\oplus \hat 0_R +\Delta(\rho),$$
with $\|\Delta(\rho)\|_1\leq \varepsilon$ and $\rho\otimes \tau\oplus
\hat 0_R\in{\cal C_Q},$ which is preserved hence completely
correctable. Let ${\cal R}$ be the TPCP map that implements the
complete correction, and let $\bar{\cal E}_p=\sum_jp_j({\cal
R}\circ{\cal E})^j$ be any convex combination.  Then
$$\bar{\cal E}(\rho\otimes \tau\oplus \hat 0_R
+\Delta(\rho))=\rho\otimes \tau\oplus \hat 0_R +\bar{\cal
E}(\Delta(\rho)).$$ 
\noindent 
Given that $\sum_jp_j({\cal R}\circ{\cal E})^j$ is a trace-norm
contraction, it follows that $|\sum_jp_j({\cal R}\circ{\cal
E})^j(\Delta(\rho))|\leq \varepsilon$.  \cvd

\subsection{Model uncertainty} 

Errors due to model uncertainty may cause the encoded information to
degrade rapidly: In fact, monotonicity of the trace norm under CPTP
dynamics is not enough to ensure non-increasing errors as in Section
\ref{encunc}.  We provide a bound on the norm-$1$ error after a finite
number of iterations of an approximately preserved code.

Since ${\cal C_Q}$ is $\varepsilon$-approximately preserved, there
exists a subsystem decomposition such that for every 
$\rho\in{\cal D(\Hi_Q)}$, we may write
$$({\cal E}\circ \Phi )(\rho)=\rho\otimes \tau\oplus \hat 0_R
+\Delta(\rho),$$
\noindent  
with $\|\Delta(\rho)\|_1\leq \varepsilon$ and ${\cal E'}|_{\cal
C_Q}={\cal E}-\Delta$ being a $1$-isometry. Choose ${\cal R}$ such
that ${\cal C_Q}$ is fixed for ${\cal R}\circ{\cal E'}|_{\cal C_Q},$
as invoked in the proof of the previous Proposition.
Therefore, \beq\label{errorsum} ({\cal R\circ E})^n(\rho)=\rho+{\cal
R}\left(\sum_{i=1}^{n}({\cal E \circ R})^{i-1}(\Delta(\rho))\right),
\eeq 
\noindent
where now $\rho \in {\cal C}_{\cal Q}$. By recalling that both ${\cal
R}$ and ${\cal E \circ R}$ act as a trace-norm contraction, it follows
that for any $i$,
$$\|{\cal R}\circ({\cal E\circ R})^i(|\Delta(\rho))\|_1\leq\|({\cal
E\circ R})^i(\Delta(\rho))\|_1 \leq \varepsilon.$$
\noindent 
We thus have the following: 

\begin{prop}\label{bound}
Any $\varepsilon$-preserved code ${\cal C_Q}$ under ${\cal E}$ admits
a correction operation $\cal R$ such that ${\cal C_Q}$ is
$\tilde{\varepsilon}$-preserved by $({\cal R\circ E})^n$, with
$\tilde{\varepsilon} \leq{n}\varepsilon$.
\end{prop}

While it is possible to construct instances in which equality holds
(at least for low $n$), an interesting question is whether the bound
can be tightened under additional information on $\Delta(\rho)$.
Suppose, for instance, that ${\cal E\circ R}$ is {\em strictly
contractive} along any trajectory ${\cal T}_\rho=\{({\cal R\circ
E})^i(\Delta(\rho));\, i \in \mathbb{N}\},\;\rho\in{\cal C_Q}$, that
is, there exists an $\alpha\in[0,1)$ (possibly dependent upon
$\varepsilon$) such that $\|{\cal E\circ R}(X)\|_1 \leq \alpha \|X\|_1
,\;\forall X\in{\cal T}_\rho$. Under these assumptions, given
Eq. \eqref{errorsum}, the relevant bound becomes a geometric sum, and
there exists a value
$$\tilde{\varepsilon}
\leq\lim_{n\rightarrow\infty}\sum_{i=1}^{n}\alpha^{i-1}\varepsilon=
\frac{\varepsilon}{1-\alpha},$$
\noindent 
so that the code is $\tilde{\varepsilon}$-correctable. The fact that
the set of strictly contractive maps is dense in the set of all CPTP
maps \cite{raginsky-strictlycontractive} leaves in principle the door
open for approximate long-term information preservation in a generic
setting. In practice, whether the error bound is acceptable for the
intended application has be tested on a case-by-case basis.  We again
provide an illustrative example.

\vspace*{1mm}

\noindent{\bf Example 2: The approximate 3-bit quantum repetition
code.} Consider the same two-qubit system and the same code described
in Example 1, but assume now that the error model is described by
$$ {\cal E}_\varepsilon=(1-\varepsilon){\cal E}+\varepsilon\,{\cal
G},\;\;\; 0<\varepsilon<1,$$
\noindent  
with ${\cal E}$ accounting for independent bit-flip errors as in
Example $1$, and ${\cal G}$ for independent phase-flip errors.  That
is, ${\cal G}\sim\{\sqrt{1-p}\, \sigma_0^{(i)},
\sqrt{\frac{p}{3}}\,\sigma_z^{(i)}\}_{i=1,2,3},$ where as before
$\sigma_z^{(i)}$ acts on the $i$-th qubit, and $\sigma_0^{(i)}=I$.

Assume that we are interested in testing how well the corrected code
performs after a given number of iterations of errors followed by
recovery, where ${\cal R}$ is chosen in the perfect case.  For
concreteness, we let $n=10$, $\varepsilon=0.050,$ and $p=0.4$ for both
${\cal E}$ and ${\cal G}$. The linear bound of Proposition \ref{bound}
predicts that the error in trace norm should be at most $\bar
\varepsilon=0.5$.  Since states represented by matrices diagonal in
the computational basis are clearly unaffected by the addition of a
dephasing term like ${\cal G}$, we choose as a test encoded state
$$\rho_c=\frac{1}{2}\begin{pmatrix}1 &1 \\1&1\end{pmatrix}.$$
After $10$ iterations, the recovered state reads  
$$\rho_{10}=\begin{pmatrix}
    0.500  &   0.332 \\
    0.332  &  0.500\end{pmatrix},$$
\noindent which leads to the following error in trace distance:
$$\varepsilon_{10}=\tr(|\rho_c-\rho_{10}|)=0.335.$$

It is easy to show that asymptotically the encoded qubit undergoes a
complete dephasing process. That is, for any initial encoded state
$\rho_0,$ we get:
$$\lim_{n\rightarrow\infty}({\cal R} \circ {\cal
E})^n(\rho_0)=\lim_{n\rightarrow\infty}({\cal R} \circ {\cal
E})^n\begin{pmatrix}a &b \\b^*&c\end{pmatrix}=\begin{pmatrix}a &0 \\0
&c\end{pmatrix}.$$ 
\noindent
Using a state like $\rho_c,$ the asymptotic error thus becomes
$\varepsilon \rightarrow 1,$ rendering this approximate code unfit for
use whenever a large number of iteration is needed.


\section{Conclusions}

We have shown that trace-norm isometries provide an operationally
motivated and mathematically consistent framework for addressing
encoding and preservation of quantum information in physical systems.
Our results point to a number of possible extensions and open
problems, some of which have been already highlighted in the main
text.  A first natural step is motivated by considering information
protection and correction in the Heisenberg picture, which calls for a
constructive characterization of observable encodings in relation to
the general subsystem structure of state encodings obtained in Sec. I.
From a general QEC standpoint, two interesting problems arise from
seeking extensions of the present $1$-isometric framework that may be
applicable to continuous-time (Markovian) dynamics and OQEC
\cite{oreshkov-continuous}, or that relax the correctability notion to
allow for non-CP transformations \cite{lidar-nonCP} (Markovian)
dynamics \cite{oreshkov-continuous}.  Since the basic mathematical
result on which our treatment relied (Busch's Theorem, from Ref.
\onlinecite{busch-isometries}) applies to arbitrary separable Hilbert
spaces in its full generality, extensions to infinite-dimensional
settings and quantum information with continuous variables are also
conceivable in principle.  Lastly, the problem of detecting when a map
is ``close'' to an isometry (with respect to the standard
Hilbert-space inner product) has recently been shown to be {\tt
QMA}-hard \cite{Rosgen2009}.  It is suggestive to speculate that
analogous complexity results for trace-norm isometries might allow to
further characterize the complexity of finding preserved quantum codes
{\tt NP}-hard \cite{viola-IPS,IPS-Long}.


\begin{acknowledgments}

We are indebted to Lajos Molnar for relevant references from the
mathematical physics literature.  F.T. acknowledges hospitality from
the Physics and Astronomy Department at Dartmouth College, where this
work was performed, and support from the University of Padova under
the QUINTET project of the Department of Information Engineering, and
the QFUTURE and CPDA080209/08 research grants.  L.V. gratefully
acknowledges partial support from the NSF under grant No. PHY05-51164.

\end{acknowledgments}

\bibliographystyle{apsrev}

\begin{thebibliography}{50}
\expandafter\ifx\csname natexlab\endcsname\relax\def\natexlab#1{#1}\fi
\expandafter\ifx\csname bibnamefont\endcsname\relax
  \def\bibnamefont#1{#1}\fi
\expandafter\ifx\csname bibfnamefont\endcsname\relax
  \def\bibfnamefont#1{#1}\fi
\expandafter\ifx\csname citenamefont\endcsname\relax
  \def\citenamefont#1{#1}\fi
\expandafter\ifx\csname url\endcsname\relax
  \def\url#1{\texttt{#1}}\fi
\expandafter\ifx\csname urlprefix\endcsname\relax\def\urlprefix{URL }\fi
\providecommand{\bibinfo}[2]{#2}
\providecommand{\eprint}[2][]{\url{#2}}

\bibitem[{\citenamefont{Viola et~al.}(2001)\citenamefont{Viola, Knill, and
  Laflamme}}]{viola-qubit}
\bibinfo{author}{\bibfnamefont{L.}~\bibnamefont{Viola}},
  \bibinfo{author}{\bibfnamefont{E.}~\bibnamefont{Knill}}, \bibnamefont{and}
  \bibinfo{author}{\bibfnamefont{R.}~\bibnamefont{Laflamme}},
  \bibinfo{journal}{J. Phys. A} {\bf 34}, {7067}
  (\bibinfo{year}{2001}).

\bibitem[{\citenamefont{Knill et~al.}(2000)\citenamefont{Knill, Laflamme, and
  Viola}}]{viola-generalnoise}
\bibinfo{author}{\bibfnamefont{E.}~\bibnamefont{Knill}},
  \bibinfo{author}{\bibfnamefont{R.}~\bibnamefont{Laflamme}}, \bibnamefont{and}
  \bibinfo{author}{\bibfnamefont{L.}~\bibnamefont{Viola}},
  \bibinfo{journal}{Phys. Rev. Lett.} \textbf{\bibinfo{volume}{84}},
  \bibinfo{pages}{2525} (\bibinfo{year}{2000}).

\bibitem[{\citenamefont{Viola and Knill}(2003)}]{verification}
\bibinfo{author}{\bibfnamefont{L.}~\bibnamefont{Viola}} \bibnamefont{and}
  \bibinfo{author}{\bibfnamefont{E.}~\bibnamefont{Knill}},
  \bibinfo{journal}{Phys. Rev. A} \textbf{\bibinfo{volume}{68}},
  \bibinfo{pages}{032311} (\bibinfo{year}{2003}).

\bibitem[{\citenamefont{Knill}(2006)}]{knill-protected}
\bibinfo{author}{\bibfnamefont{E.}~\bibnamefont{Knill}},
  \bibinfo{journal}{Phys. Rev. A} \textbf{\bibinfo{volume}{74}},
  \bibinfo{pages}{042301} (\bibinfo{year}{2006}).

\bibitem[{\citenamefont{Zanardi and Rasetti}(1997)}]{zanardi-DFS}
\bibinfo{author}{\bibfnamefont{P.}~\bibnamefont{Zanardi}} \bibnamefont{and}
  \bibinfo{author}{\bibfnamefont{M.}~\bibnamefont{Rasetti}},
  \bibinfo{journal}{Phys. Rev. Lett.} \textbf{\bibinfo{volume}{79}},
  \bibinfo{pages}{3306} (\bibinfo{year}{1997}).

\bibitem[{\citenamefont{Lidar et~al.}(1997)\citenamefont{Lidar, Chuang, and
  Whaley}}]{lidar-DFS}
\bibinfo{author}{\bibfnamefont{D.~A.} \bibnamefont{Lidar}},
  \bibinfo{author}{\bibfnamefont{I.~L.} \bibnamefont{Chuang}},
  \bibnamefont{and} \bibinfo{author}{\bibfnamefont{K.~B.}
  \bibnamefont{Whaley}}, \bibinfo{journal}{Phys. Rev. Lett.}
  \textbf{\bibinfo{volume}{81}}, \bibinfo{pages}{2594} (\bibinfo{year}{1997}).

\bibitem[{\citenamefont{Kribs et~al.}(2005{\natexlab{a}})\citenamefont{Kribs,
  Laflamme, and Poulin}}]{kribs-QEC}
\bibinfo{author}{\bibfnamefont{D.~W.} \bibnamefont{Kribs}},
  \bibinfo{author}{\bibfnamefont{R.}~\bibnamefont{Laflamme}}, \bibnamefont{and}
  \bibinfo{author}{\bibfnamefont{D.}~\bibnamefont{Poulin}},
  \bibinfo{journal}{Phys. Rev. Lett.} \textbf{\bibinfo{volume}{94}},
  \bibinfo{pages}{180501} (\bibinfo{year}{2005}{\natexlab{a}}).

\bibitem[{\citenamefont{Kribs et~al.}(2006)\citenamefont{Kribs, Laflamme,
  Poulin, and Lesosky}}]{kribs-OEC}
\bibinfo{author}{\bibfnamefont{D.~W.} \bibnamefont{Kribs}},
  \bibinfo{author}{\bibfnamefont{R.}~\bibnamefont{Laflamme}},
  \bibinfo{author}{\bibfnamefont{D.}~\bibnamefont{Poulin}}, \bibnamefont{and}
  \bibinfo{author}{\bibfnamefont{M.}~\bibnamefont{Lesosky}},
  \bibinfo{journal}{Quantum Inf. Comput.}
  \textbf{\bibinfo{volume}{6}}, \bibinfo{pages}{382} (\bibinfo{year}{2006}).

\bibitem[{\citenamefont{Brun et~al.}(2006)\citenamefont{Brun, Devetak, and
  Hsieh}}]{Brun-EAQEC}
\bibinfo{author}{\bibfnamefont{T.}~\bibnamefont{Brun}},
  \bibinfo{author}{\bibfnamefont{I.}~\bibnamefont{Devetak}}, \bibnamefont{and}
  \bibinfo{author}{\bibfnamefont{M.~H.} \bibnamefont{Hsieh}},
  \bibinfo{journal}{Science} \textbf{\bibinfo{volume}{314}},
  \bibinfo{pages}{436} (\bibinfo{year}{2006}).

\bibitem[{\citenamefont{Blume-Kohout et~al.}(2008)\citenamefont{Blume-Kohout,
  Ng, Poulin, and Viola}}]{viola-IPS}
\bibinfo{author}{\bibfnamefont{R.}~\bibnamefont{Blume-Kohout}},
  \bibinfo{author}{\bibfnamefont{H.~K.} \bibnamefont{Ng}},
  \bibinfo{author}{\bibfnamefont{D.}~\bibnamefont{Poulin}}, \bibnamefont{and}
  \bibinfo{author}{\bibfnamefont{L.}~\bibnamefont{Viola}},
  \bibinfo{journal}{Phys. Rev. Lett.} \textbf{\bibinfo{volume}{100}},
  \bibinfo{pages}{030501} (\bibinfo{year}{2008}).

\bibitem[{\citenamefont{Blume-Kohout et~al.}(2009)\citenamefont{Blume-Kohout,
  Ng, Poulin, and Viola}}]{IPS-Long}
\bibinfo{author}{\bibfnamefont{R.}~\bibnamefont{Blume-Kohout}},
  \bibinfo{author}{\bibfnamefont{H.~K.} \bibnamefont{Ng}},
  \bibinfo{author}{\bibfnamefont{D.}~\bibnamefont{Poulin}}, \bibnamefont{and}
  \bibinfo{author}{\bibfnamefont{L.}~\bibnamefont{Viola}}, \bibinfo{journal}{in
  preparation}  (\bibinfo{year}{2009}).

\bibitem[{\citenamefont{Fuchs and van~de Graaf}(1999)}]{fuchs-distance}
\bibinfo{author}{\bibfnamefont{C.~A.} \bibnamefont{Fuchs}} \bibnamefont{and}
  \bibinfo{author}{\bibfnamefont{J.}~\bibnamefont{van~de Graaf}},
  \bibinfo{journal}{IEEE Trans. Inf. Theory} \textbf{\bibinfo{volume}{45}},
  \bibinfo{pages}{1216} (\bibinfo{year}{1999}).

\bibitem[{\citenamefont{Lidar et~al.}(2008)\citenamefont{Lidar, Zanardi, and
  Khodjasteh}}]{lidar-distance}
\bibinfo{author}{\bibfnamefont{D.~A.} \bibnamefont{Lidar}},
  \bibinfo{author}{\bibfnamefont{P.}~\bibnamefont{Zanardi}}, \bibnamefont{and}
  \bibinfo{author}{\bibfnamefont{K.}~\bibnamefont{Khodjasteh}},
  \bibinfo{journal}{Phys. Rev. A} \textbf{\bibinfo{volume}{78}},
  \bibinfo{pages}{012308} (\bibinfo{year}{2008}).

\bibitem[{\citenamefont{Khodjsateh et~al.}(2009)\citenamefont{Khodjsateh,
  Lidar, and Viola}}]{CDCG}
\bibinfo{author}{\bibfnamefont{K.}~\bibnamefont{Khodjasteh}},
  \bibinfo{author}{\bibfnamefont{D.~A.} \bibnamefont{Lidar}}, \bibnamefont{and}
  \bibinfo{author}{\bibfnamefont{L.}~\bibnamefont{Viola}},
  \bibinfo{journal}{arXiv: 0908.1526}  (\bibinfo{year}{2009}).

\bibitem[{\citenamefont{Kribs and Spekkens}(2006)}]{spekkens-unitary}
\bibinfo{author}{\bibfnamefont{D.~W.} \bibnamefont{Kribs}} \bibnamefont{and}
  \bibinfo{author}{\bibfnamefont{R.~W.} \bibnamefont{Spekkens}},
  \bibinfo{journal}{Phys. Rev. A}
  \textbf{\bibinfo{volume}{74}}, \bibinfo{pages}{042329}
  (\bibinfo{year}{2006}).

\bibitem[{\citenamefont{Choi et~al.}(2009)\citenamefont{Choi, Johnston, and
  Kribs}}]{choi-multiplicative}
\bibinfo{author}{\bibfnamefont{M.-D.} \bibnamefont{Choi}},
  \bibinfo{author}{\bibfnamefont{N.}~\bibnamefont{Johnston}}, \bibnamefont{and}
  \bibinfo{author}{\bibfnamefont{D.~W.} \bibnamefont{Kribs}},
  \bibinfo{journal}{J Phys. A}
  \textbf{\bibinfo{volume}{42}}, \bibinfo{pages}{245303}
  (\bibinfo{year}{2009}).

\bibitem[{\citenamefont{Schumacher and Westmoreland}(2009)}]{Tyson}
\bibinfo{author}{\bibfnamefont{B.}~\bibnamefont{Schumacher}} \bibnamefont{and}
  \bibinfo{author}{\bibfnamefont{M.~D.} \bibnamefont{Westmoreland}},
  \bibinfo{journal}{J. Math. Phys.} \textbf{\bibinfo{volume}{50}},
  \bibinfo{pages}{032106} (\bibinfo{year}{2009}).

\bibitem[{\citenamefont{B\'eny and Oreshkov}(2009)}]{beny-approximate}
\bibinfo{author}{\bibfnamefont{C.}~\bibnamefont{B\'eny}} \bibnamefont{and}
  \bibinfo{author}{\bibfnamefont{O.}~\bibnamefont{Oreshkov}},
  \bibinfo{journal}{arXiv:0907.5391}  (\bibinfo{year}{2009}).

\bibitem[{\citenamefont{B\'eny}(2009)}]{beny-alg}
\bibinfo{author}{\bibfnamefont{C.}~\bibnamefont{B\'eny}},
  \bibinfo{journal}{arXiv:0907.4207}  (\bibinfo{year}{2009}).

\bibitem[{\citenamefont{Ng and Mandayam}(2009)}]{Ng2009}
\bibinfo{author}{\bibfnamefont{H.~K.} \bibnamefont{Ng}} \bibnamefont{and}
  \bibinfo{author}{\bibfnamefont{P.}~\bibnamefont{Mandayam}},
  \bibinfo{journal}{arXiv:0909.0931}  (\bibinfo{year}{2009}).

\bibitem[{Rem({\natexlab{a}})}]{Remark1}
\bibinfo{note}{Note that the domain of $\Phi$ is extended to arbitrary positive
  semidefinite operators by defining: $\Phi(\Pi_A^\lambda \rho
  \Pi_A^\lambda)=\tr(\Pi_A^\lambda \rho) \Phi\left(\Pi_A^\lambda \rho
  \Pi_A^\lambda/ \tr(\Pi_A^\lambda \rho) \right)$.}

\bibitem[{\citenamefont{Holevo}(2001)}]{holevo}
\bibinfo{author}{\bibfnamefont{A.}~\bibnamefont{Holevo}},
  \emph{\bibinfo{title}{Statistical Structure of Quantum Theory}}, Lecture
  Notes in Physics; Monographs: {\bf 67} (\bibinfo{publisher}{Springer-Verlag,
  Berlin}, \bibinfo{year}{2001}).

\bibitem[{\citenamefont{Petz}(2008)}]{petz-qstatistics}
\bibinfo{author}{\bibfnamefont{D.}~\bibnamefont{Petz}},
  \emph{\bibinfo{title}{Quantum Information Theory and Quantum Statistics}}
  (\bibinfo{publisher}{Springer-Verlag, Berlin Heidelberg},
  \bibinfo{year}{2008}).

\bibitem[{\citenamefont{Nielsen and Chuang}(2002)}]{nielsen-chuang}
\bibinfo{author}{\bibfnamefont{M.~A.} \bibnamefont{Nielsen}} \bibnamefont{and}
  \bibinfo{author}{\bibfnamefont{I.~L.} \bibnamefont{Chuang}},
  \emph{\bibinfo{title}{Quantum Computation and Information}}
  (\bibinfo{publisher}{Cambridge University Press, Cambridge},
  \bibinfo{year}{2002}).

\bibitem[{\citenamefont{Busch}(1999)}]{busch-isometries}
\bibinfo{author}{\bibfnamefont{P.}~\bibnamefont{Busch}},
  \bibinfo{journal}{Math. Phys. Anal. Geom.}
  \textbf{\bibinfo{volume}{2}}, \bibinfo{pages}{83} (\bibinfo{year}{1999}).

\bibitem[{\citenamefont{Molnar and Timmermann}(2003)}]{molnar-isometries}
\bibinfo{author}{\bibfnamefont{L.}~\bibnamefont{Molnar}} \bibnamefont{and}
  \bibinfo{author}{\bibfnamefont{W.}~\bibnamefont{Timmermann}},
  \bibinfo{journal}{J. Phys. A} \textbf{\bibinfo{volume}{36}},
  \bibinfo{pages}{267} (\bibinfo{year}{2003}).

\bibitem[{Rem({\natexlab{b}})}]{Remark2}
\bibinfo{note}{A related result by P. Mankiewicz [Thm. 5 in Bull. Acad. Polon.
  Sciences {\bf 20}, 367 (1972)] states that any bijective isometry between
  convex sets with {\em nonempty interiors} can be uniquely extended to a
  bijective affine isometry between the whole spaces. Note that a direct
  application of this result to $1$-isometric (not necessarily linear)
  encodings defined on density ${\mathcal D}({\mathcal H}_Q))$ is prevented by
  the fact that the image $\Phi({\mathcal D}({\mathcal H}_Q))$ need not have
  nonempty interior in general.}

\bibitem[{Rem({\natexlab{d}})}]{Remark3} \bibinfo{note}{In principle,
different choices of observable encodings may be considered. For
example, $I_F$ may be substituted by a different fixed operator $X_F$,
even if in that case some of the faithfulness requirements may need
reconsideration as they may fail. In practice, beside being directly
suggested by the results on faithful encodings, the choice of $I_F$ on
$\Hi_F$ is particularly convenient, as it also ensures robustness with
respect to the co-factor state (see Sec. \ref{subsystems}). Also, the
choice of $X_R$ could in principle depend on $A,$ but it would
nevertheless remain irrelevant with respect to the faithfulness
requirements, given the form of an isometric encoding.}

\bibitem[{\citenamefont{Kribs et~al.}(2005{\natexlab{b}})\citenamefont{Kribs,
  Laflamme, and Poulin}}]{kribs-oqec}
\bibinfo{author}{\bibfnamefont{D.}~\bibnamefont{Kribs}},
  \bibinfo{author}{\bibfnamefont{R.}~\bibnamefont{Laflamme}}, \bibnamefont{and}
  \bibinfo{author}{\bibfnamefont{D.}~\bibnamefont{Poulin}},
  \bibinfo{journal}{Phys. Rev. Lett.} \textbf{\bibinfo{volume}{94}},
  \bibinfo{pages}{180501} (\bibinfo{year}{2005}{\natexlab{b}}).

\bibitem[{\citenamefont{Lindblad}(1999)}]{lindblad-nocloning}
\bibinfo{author}{\bibfnamefont{G.}~\bibnamefont{Lindblad}},
  \bibinfo{journal}{Lett. Math. Phys.}
  \textbf{\bibinfo{volume}{47}}, \bibinfo{pages}{189} (\bibinfo{year}{1999}).

\bibitem[{\citenamefont{Knill and Laflamme}(1997)}]{knill-QEC}
\bibinfo{author}{\bibfnamefont{E.}~\bibnamefont{Knill}} \bibnamefont{and}
  \bibinfo{author}{\bibfnamefont{R.}~\bibnamefont{Laflamme}},
  \bibinfo{journal}{Phys. Rev. A}
  \textbf{\bibinfo{volume}{55}}, \bibinfo{pages}{900} (\bibinfo{year}{1997}).

\bibitem[{\citenamefont{Bacon}(2006)}]{Bacon}
\bibinfo{author}{\bibfnamefont{D.}~\bibnamefont{Bacon}},
  \bibinfo{journal}{Phys. Rev. A} \textbf{\bibinfo{volume}{73}},
  \bibinfo{pages}{012340} (\bibinfo{year}{2006}).

\bibitem[{\citenamefont{Barnum and Knill}(2002)}]{barnum-reversing}
\bibinfo{author}{\bibfnamefont{H.}~\bibnamefont{Barnum}} \bibnamefont{and}
  \bibinfo{author}{\bibfnamefont{E.}~\bibnamefont{Knill}},
  \bibinfo{journal}{J. Math. Phys}
  \textbf{\bibinfo{volume}{43}}, \bibinfo{pages}{2097} (\bibinfo{year}{2002}).

\bibitem[{\citenamefont{Ticozzi and Pavon}(2008)}]{ticozzi-htheorem}
\bibinfo{author}{\bibfnamefont{F.}~\bibnamefont{Ticozzi}} \bibnamefont{and}
  \bibinfo{author}{\bibfnamefont{M.}~\bibnamefont{Pavon}},
  \bibinfo{journal}{arXiv:0811.0929}  (\bibinfo{year}{2008}).

\bibitem[{\citenamefont{Schumacher and
  Westmoreland}(2002)}]{schumacher-approximate}
\bibinfo{author}{\bibfnamefont{B.}~\bibnamefont{Schumacher}} \bibnamefont{and}
  \bibinfo{author}{\bibfnamefont{M.~D.} \bibnamefont{Westmoreland}},
  \bibinfo{journal}{Quantum Inf. Process.}
  \textbf{\bibinfo{volume}{1}}, \bibinfo{pages}{5} (\bibinfo{year}{2002}).

\bibitem[{\citenamefont{Klesse}(2007)}]{klesse-approximate}
\bibinfo{author}{\bibfnamefont{R.}~\bibnamefont{Klesse}},
  \bibinfo{journal}{Phys. Rev. A} \textbf{\bibinfo{volume}{75}},
  \bibinfo{pages}{062315} (\bibinfo{year}{2007}).

\bibitem[{\citenamefont{Buscemi}(2008)}]{buscemi-approximate}
\bibinfo{author}{\bibfnamefont{F.}~\bibnamefont{Buscemi}},
  \bibinfo{journal}{Phys. Rev. A}
  \textbf{\bibinfo{volume}{77}}, \bibinfo{pages}{012309}
  (\bibinfo{year}{2008}).

\bibitem[{\citenamefont{Leung et~al.}(1997)\citenamefont{Leung, A., Nielsen,
  Chuang, and Yamamoto}}]{leung-approximate}
\bibinfo{author}{\bibfnamefont{D.~W.} \bibnamefont{Leung}},
  \bibinfo{author}{\bibfnamefont{M.}~\bibnamefont{A.}}
  \bibinfo{author}{\bibnamefont{Nielsen}},
  \bibinfo{author}{\bibfnamefont{I.~L.} \bibnamefont{Chuang}},
  \bibnamefont{and} \bibinfo{author}{\bibfnamefont{Y.}~\bibnamefont{Yamamoto}},
  \bibinfo{journal}{Phys. Rev. A} \textbf{\bibinfo{volume}{56}},
  \bibinfo{pages}{2567} (\bibinfo{year}{1997}).

\bibitem[{\citenamefont{Hyers and Ulam}(1945)}]{hyers-ulam}
\bibinfo{author}{\bibfnamefont{D.~H.} \bibnamefont{Hyers}} \bibnamefont{and}
  \bibinfo{author}{\bibfnamefont{S.~M.} \bibnamefont{Ulam}},
  \bibinfo{journal}{Bull. Am. Math. Soc.}
  \textbf{\bibinfo{volume}{51}}, \bibinfo{pages}{288} (\bibinfo{year}{1945}).

\bibitem[{Rem({\natexlab{e}})}]{Remark4} \bibinfo{note}{Let $\cal A,B$
be Banach spaces. A map $f:{\cal A}\rightarrow {\cal B}$ is an {\em
$\varepsilon$-isometry} if for every $x,y \in{\cal A}$ it obeys 
$\left| \|f(x)-f(y)\|-\|x-y\| \right|\leq \varepsilon$, for some
$\varepsilon\in {\mathbb R}^+$.}

\bibitem[{\citenamefont{Bhatia and Semrl}(1997)}]{bhatia-semrl}
\bibinfo{author}{\bibfnamefont{R.}~\bibnamefont{Bhatia}} \bibnamefont{and}
  \bibinfo{author}{\bibfnamefont{P.}~\bibnamefont{Semrl}},
  \bibinfo{journal}{Am. Math. Month.}
  \textbf{\bibinfo{volume}{104}}, \bibinfo{pages}{497} (\bibinfo{year}{1997}).

\bibitem[{\citenamefont{Dilworth}(1999)}]{dilworth}
\bibinfo{author}{\bibfnamefont{S.~J.} \bibnamefont{Dilworth}},
  \bibinfo{journal}{Bull. Lon. Math. Soc.}
  \textbf{\bibinfo{volume}{31}}, \bibinfo{pages}{471} (\bibinfo{year}{1999}).

\bibitem[{\citenamefont{Rassias}(2001)}]{rassias}
\bibinfo{author}{\bibfnamefont{T.}~\bibnamefont{Rassias}},
  \bibinfo{journal}{Internat. J. Math. Math. Sci}
  \textbf{\bibinfo{volume}{25}}, \bibinfo{pages}{73} (\bibinfo{year}{2001}).

\bibitem[{\citenamefont{Liu and Zhang}(2009)}]{liu-approximate}
\bibinfo{author}{\bibfnamefont{R.}~\bibnamefont{Liu}} \bibnamefont{and}
  \bibinfo{author}{\bibfnamefont{L.}~\bibnamefont{Zhang}},
  \bibinfo{journal}{J. Math. Anal. App.}
  \textbf{\bibinfo{volume}{352}}, \bibinfo{pages}{749} (\bibinfo{year}{2009}).

\bibitem[{\citenamefont{Raginsky}(2002)}]{raginsky-strictlycontractive}
\bibinfo{author}{\bibfnamefont{M.}~\bibnamefont{Raginsky}},
  \bibinfo{journal}{Phys. Rev. A} \textbf{\bibinfo{volume}{65}},
  \bibinfo{pages}{032306} (\bibinfo{year}{2002}).

\bibitem[{\citenamefont{Oreshkov et~al.}(2008)\citenamefont{Oreshkov, Lidar,
  and Brun}}]{oreshkov-continuous}
\bibinfo{author}{\bibfnamefont{O.}~\bibnamefont{Oreshkov}},
  \bibinfo{author}{\bibfnamefont{D.~A.} \bibnamefont{Lidar}}, \bibnamefont{and}
  \bibinfo{author}{\bibfnamefont{T.}~\bibnamefont{Brun}},
  \bibinfo{journal}{arXiv:0806.3145}  (\bibinfo{year}{2008}).

\bibitem[{\citenamefont{Shabani and Lidar}(2009)}]{lidar-nonCP}
\bibinfo{author}{\bibfnamefont{A.}~\bibnamefont{Shabani}} \bibnamefont{and}
  \bibinfo{author}{\bibfnamefont{D.~A.} \bibnamefont{Lidar}},
  \bibinfo{journal}{Phys. Rev. A} \textbf{\bibinfo{volume}{80}},
  \bibinfo{pages}{012309} (\bibinfo{year}{2009}).

\bibitem[{\citenamefont{Rosgen}(2009)}]{Rosgen2009}
\bibinfo{author}{\bibfnamefont{B.}~\bibnamefont{Rosgen}},
  \bibinfo{journal}{arXiv:0910.3740}  (\bibinfo{year}{2009}).


\end{thebibliography}

\end{document}